\documentclass[twocolumn,pre,showpacs,superscriptaddress]{revtex4}
\usepackage{graphics,epsfig}

\begin{document}

\title{Dynamic instabilities induced by asymmetric
    influence: Prisoners' dilemma game on small-world networks}
\author {Beom Jun \surname{Kim}}
\email{beomjun@ajou.ac.kr}
\affiliation{Department of Molecular Science
  and Technology, Ajou University, Suwon 442-749, Korea}
\author {Ala \surname{Trusina}}
\author {Petter \surname{Holme}}
\author {Petter \surname{Minnhagen}}
\affiliation {Department of Theoretical Physics,
Ume{\aa} University, 901 87 Ume{\aa}, Sweden }
\author{Jean S. \surname{Chung}}
\affiliation{Department of Physics, Chungbuk National University,
  Cheongju 361-763, Korea}
\author{M.Y. \surname{Choi}}
\affiliation{Department of Physics and Center for Theoretical Physics,
Seoul National University, Seoul 151-747, Korea}

\begin{abstract}
A two-dimensional small-world type network, subject to spatial
prisoners' dilemma dynamics and containing an influential
node defined as a special node with a finite density of directed
random links to the other nodes in the network, is numerically
investigated. It is shown that the degree of
cooperation does not remain at a steady state level but displays a
punctuated equilibrium type behavior manifested by the existence of sudden breakdowns
of cooperation. The breakdown of cooperation is linked to an imitation
of a successful selfish strategy of the influential node.
It is also found that while the breakdown of cooperation 
occurs suddenly, the recovery of it requires longer time. This recovery
time may, depending on the degree of steady state cooperation,
either increase or decrease with an increasing number of long
range connections.
\end{abstract}

\pacs{87.23.Kg, 84.35.+i, 87.23.Ge, 02.50.Le}

\maketitle
\section{Introduction}
\begin{figure}
\centering{\resizebox*{!}{4.3cm}{\includegraphics{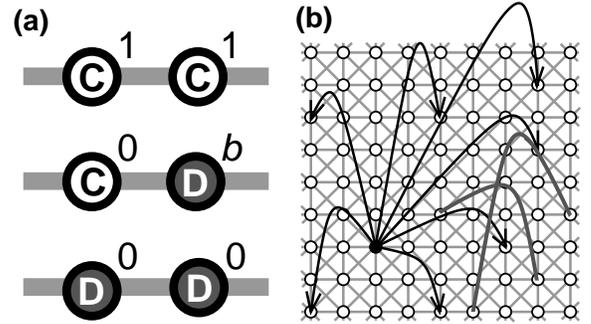}}}
\caption{ 
(a) The encounter payoff: When two cooperators (C) encounter, both
  score unity. When a cooperator meets a defector
  (D) the defector score $b$ and the cooperator 0. An encounter between two
  defectors results in 0 for both nodes.
(b) The network: A two-dimensional square lattice with eight
  nearest neighbors and long range ``short-cuts''
  are randomly added (red lines without arrows).
  The influential node (starting point for lines with arrows)
  effects the network over long ranges through unidirectional connections
  (lines with arrows). 
}  
\label{fig:network}
\end{figure}

Ever since it's introduction iterated Prisoners' Dilemma games has
been central in understanding the conditions for cooperation among
populations of selfish individuals.~\cite{axelrod}
Applications has ranged from RNA virus interactions~\cite{virus} to
Westernization in central Africa~\cite{africa}, and consequently
a variety of generalizations has been studied.
The present work takes the spatial Prisoners' Dilemma of Nowak
\textit{et al.}~\cite{nowak-all} as its starting-point.~\cite{other_spatial}
Here the players
are situated on a two-dimensional lattice, interacting only with their
neighbors. Rather than examining the stability of strategies based
on memory of the opponent's behavior, as in the ordinary iterated
Prisoners' Dilemma, the spatial Prisoners' Dilemma serves to answer
questions such as under what conditions cooperation can be stable in
(social) space.~\cite{spd} Following Refs.~\cite{nowak-all} the interactions
can be chosen as simple as follows: The payoff is simultaneously
calculated for every node (player). The
contribution to the gain from an encounter is illustrated in
Fig.~\ref{fig:network}(a); the sum of the encounters from each neighbor
gives the gain for a certain node.  In the next move each node follows
the most successful neighbor. (This is a feature of
successful strategies such as tit-for-tat~\cite{axelrod} or win-stay
lose-shift~\cite{WSLS} of the two-player Prisoners' Dilemma.)
Defined in this way, the dynamics may e.g.\
reflect that of groups of individuals with mutual trust and cooperation
interacting with social regions of unrest. To add the element of
occasional irrational moves by individuals, and get a way from a
purely deterministic dynamics, one can allow for `mutations': a random
strategy (D or C is chosen randomly) is assigned to a player with probability
$p_m$.

Important features of social networks such as high clustering and
short characteristic path-length can be modeled by the
Watts and Strogatz (WS) model~\cite{WS,watts}, where the links of a
regular network are randomly rewired to introduce long-range ``short-cuts''.
On a one-dimensional small-world network the presence of 
long-range connections has been found to increase the 
density of defectors~\cite{abramson}.
To get closer to the original work by Nowak \textit{et al.} we start from
a two-dimensional WS model network. In society, massmedial persons may influence
others much stronger than the average individual, still these influential
persons are coupled back to their social surroundings. One concrete
example along this general line
is smoking among adolescents, a behavior spurred by both the
individual's social surroundings and role models of the media.~\cite{unger}
To model this
situation we let one node have additional directed links randomly
distributed outwards to the rest of the network. In this way we hope to
catch some general effects that such an influential node might have on 
the dynamical behavior of a social network.

\section{The Model}
The starting point is a $L\times L$ square grid
(with periodic boundary conditions)
where each node has eight neighbors reachable by a chess king's move.
Long range bidirectional links are added with a
probability $p$ making the average number of short-cuts $Np$ ($N=L^2$).
One node is randomly chosen as the influential node and in addition
to its local bidirectional connections, this
node is unidirectionally connected to arbitrary nodes of the network with
a probability $p_s$.
These additional links are directed so that nodes unidirectionally
connected to the special node sees the special node as one of its
neighbors, but not vice versa.
The influential node only gets feedback from its local
mutual connections. (See Fig.~\ref{fig:network}(b).)

In our simulations we use a typical
lattice size is $L=32$, with the number of additional directed
connections to the influential node given by $Np_s$ with $p_s$
typically 0.2, the mutation rate $p_m$ typically 0.001, the
shortcut density $p$ from 0 to 0.1, and $\mathcal{O}(100)$ network
realizations. The gain of the certain node
(in our version of prisoners' dilemma game)
is calculated as the average score of the individual encounters:
the sum of the encounters from each neighbor is divided by the number of
the neighbors. This normalization is done to avoid an additional bias
from the higher degree of some nodes, and thus keep the game closer Nowak
and May's original spatial prisoners' dilemma game.

\begin{figure}
\centering{\resizebox*{!}{5.3cm}{\rotatebox{270}{\includegraphics{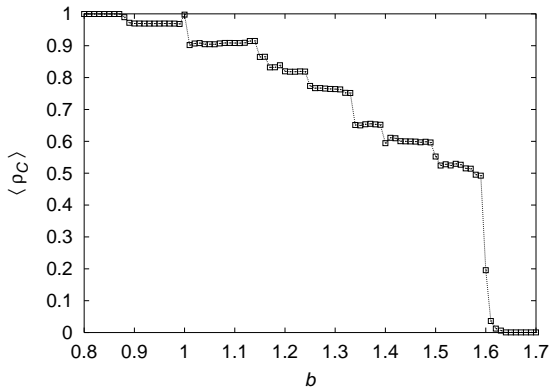}}}}
\caption{The averaged cooperator density in a regular network with an
  influential node versus temptation $b$. For $7/8<b<8/5$ we have
  $0<\langle\rho_c\rangle< 1$. The two cases we study the time evolution for
  is $b=1.3$ and $b=1.45$.
  }
\label{steps}
\end{figure}

\section{Simulation Results}
In order to analyze the dynamics of this model we start by calculating the average
density of cooperators $\rho_C$ as a function of the pay-off $b$
between defector D and cooperator C (see Fig.~\ref{fig:network}(a)).
As seen in Fig.~\ref{steps} $\rho_C$
has a step structure. These steps reflect the interplay between the
underlying spatial structure and the PD dynamics~\cite{nowak-all}:
Each level is
characterized by the condition that $n$ C's wins over $m$ D's and
consequently the step condition given by $n=bm$ and the sequence of steps
discernible in Fig.~\ref{steps} is  7/8, 1, 8/7, 7/6, 6/5, 5/4, 4/3, 7/5,
3/2, 8/5 corresponding to the case
when $p_s=0$ and the additional steps at 8/9, 9/8 due to the
additional coupling for nodes attached to the influential node. For
$b>8/5$ there is no cooperation left and $\rho_C=0$ and for $b<7/8$
cooperation wins and $\rho_c=1$.

\begin{figure}
\centering{\resizebox*{!}{5.3cm}{\rotatebox{270}{\includegraphics{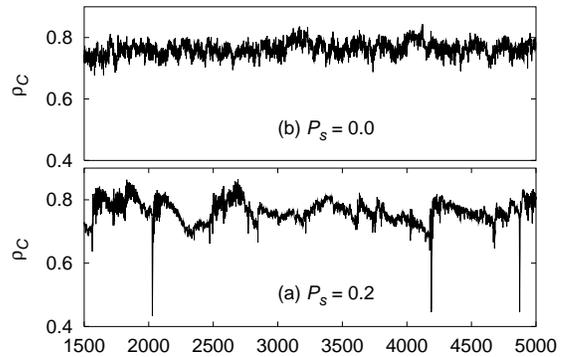}}}}
\caption{The time evolution of cooperator density. Without (a) and
  with (b) ``influential node'' node. The temptation is $b=1.3$.
  }
\label{fig:time}
\end{figure}

\begin{figure}
  \centering{\resizebox*{!}{5.3cm}{\rotatebox{270}{\includegraphics{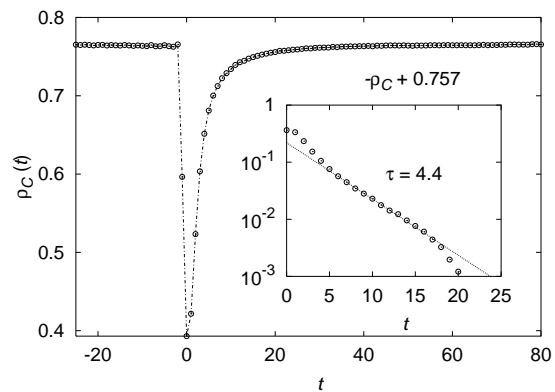}}}}
  \caption{The jump structure obtained from the average over about thousand
    jumps in Fig.~\ref{fig:time}. The sharp decrease of cooperator density
    $\rho_C$ is followed by a gradual recovery to the equilibrium value.
    Inset: The
    long-time recovery behavior is well described by an exponential 
    $|\rho_C-\langle\rho_C\rangle|\propto \exp (-t/ \tau)$ with the recovery
    time $\tau\approx 4.4$.
    }
  \label{fig:reltime}
\end{figure}

In the following we will focus on $b=1.3$ which is associated with a
plateau in the middle with $\rho_C\approx 0.76$.
Fig.~\ref{fig:time}(a) shows the
time evolution for $b=1.3$ and $p_s=0$ i.e.\ the case when there
is no influential node. In this case the level of cooperation remains
stable with relatively small fluctuations around the average value.
This feature is dramatically changed when we introduce the special
influential node as shown in Fig.~\ref{fig:time}(b)
for $p_s=0.2$. The equilibrium is
now punctuated by sudden drops of cooperation.
In Fig.~\ref{fig:reltime} we display the
average drop (obtained by averaging over about thousand sudden drops).
The typical feature is a very dramatic sudden jump followed by a slower
recovering back to the steady state situation. This recovery
back to steady state is exponential as demonstrated
in the inset of Fig.~\ref{fig:reltime}.

\begin{figure*}
\resizebox*{17.5cm}{!}{\includegraphics{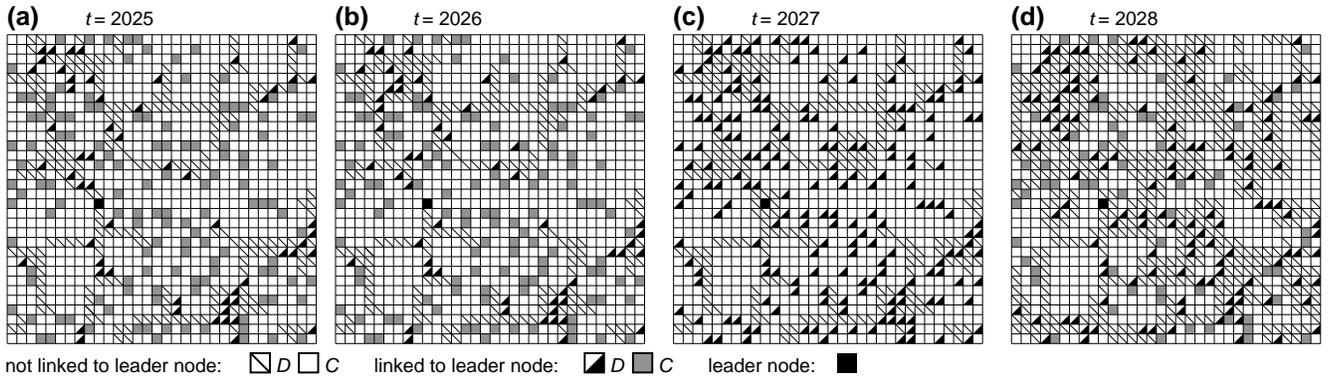}}
\caption{Complete network configuration at the four consecutive time steps of
  the run illustrated in Fig.~\ref{fig:time}: In (a) the gain of the leader
  node (that is a defector) scores $5b/8$, in (b) the score of the  leader node
  increases to $7b/8$ and in (c) the defecting strategy
  spreads through the directed links, and further on to the surrounding of the end
  nodes of the directed links (d). ``Linked to'' in the legend means ``having a
  directed link from the leader node.''
  }
\label{fig:evol}
\end{figure*}

As a first step we investigate what exactly
triggers the sudden drop of cooperation: The basic mechanisms is
that a situation arises where the influential node as a
defector gets a very high score. The successful defector strategy of
the influential node is then rapidly
spread through the directed links from this node
i.e.\ the sudden drop in cooperation is triggered by an imitation of a
successful selfish behavior of the influential node.
Figure~\ref{fig:evol}
shows a typical example of how the triggering high score situation is built up
in the environment of the influential node. The figure shows four
consecutive time steps for the same run as in Fig.~\ref{fig:time}. In
the second timestep (Fig.~\ref{fig:evol}(b)) the influential node
is surrounded by seven cooperators and hence gets the high score $7b/8$.
This high score causes an instability since it causes the
defector strategy to be imitated both by the immediate surrounding and by the rest of the
network through the directed links from the influential nodes (Fig.~\ref{fig:evol}(c)).
In the next step (Fig.~\ref{fig:evol}(d)) the defector strategy
spreads to the nodes in the vicinity of the nodes connected to the
influential node.

\begin{figure}
  \centering{\resizebox*{7.4cm}{!}{{\includegraphics{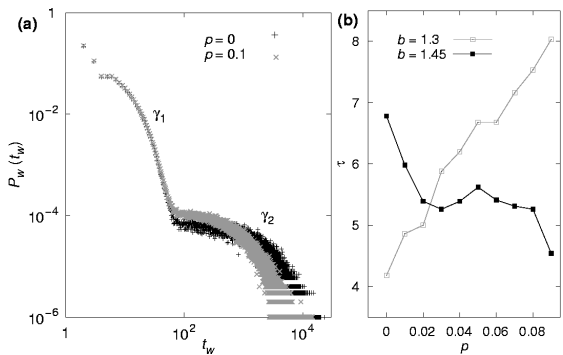}}}}
  \caption{(a) Averaged probability distribution $P_w(t_w)$ of the
    waiting time $t_w$ (time between breakdowns) for $b=1.3$, $p=0.1$
    and $p_m=0.001$. This distribution to good approximation consists
    of two exponential parts
  $\propto \exp(-x/\gamma)$ with the time scales $\gamma_1=8.0 \pm 0.1$,
  $\gamma_2=993 \pm 7$, respectively. Without short cuts ($p=0$) the
  time scales are $\gamma_1= 7.9 \pm 0.1$, $\gamma_2= 1945 \pm 4$.
  Thus the effect of adding short cuts basically just speeds up the time
  evolution. (b) The recovery time $\tau$ (see  Fig.~\ref{fig:reltime}) versus
small world
rewiring probability $p$ at two different temptations: $b=1.3$ and $b=1.45$.
The recovery time decreases with
increasing number of long-range connections in case of $b=1.3$ and
increases  for $b=1.45$. Consequently, long range connection can effect the recovering back to steady state in opposite ways depending on the steady state proportion between defectors and cooperators.
}
\label{wtdistr_p0}
\end{figure}

How often does such a breakdown occur?
Figure~\ref{wtdistr_p0} shows the average
probability distribution for the waiting time between two breakdowns.
The waiting time
distribution $P_w(t_w)$ is clearly exponential
for large $t_w$. In addition it has some structure as discussed below.

In order to gain some further insight we investigate how the recovery
time and waiting time depends on the parameters of the model.
The waiting time distribution does not change qualitatively when a 
rewiring probability is introduced. The only change is a small
quantitative decrease in the average recovery time. This is in accord
with the intuitive idea that more long range connection will in general speed up
the time evolution. In our particular model it means that the triggering type
situation (shown in Fig.~\ref{fig:evol}(b)) will
arise more frequently when long range
connections are present. The structure of the waiting time
distribution consists to good approximation of two exponential
decays as shown in in Fig.~\ref{wtdistr_p0} (a).
This structure of the waiting time distribution is caused by an interplay between the spatial lattice and the
PD pay-off.

Figure~\ref{wtdistr_p0}(b)
shows how the recovery time $\tau$ depends on the
rewiring probability $p$. The striking thing here is that for $b=1.3$ and
$\rho_c\approx0.76$
the recovery time increases with
increasing $p$ so that actually more connections between different
parts of the network will slow down the recovery.
However for $b=1.45$ and $\rho_c\approx 0.6$ the recovery time
instead decreases with increasing $p$ as also shown in Fig.~\ref{wtdistr_p0}(b)
Consequently the change in the recovery time with $p$ depends on the relative
proportion of defectors and collaborators in the steady state situation:
If the cooperator density is large enough then an additional short-cut
will more often connect a defector to a cooperator which promotes the
defector strategy and slows down the recovery. If the cooperator
density is smaller the situation changes and an increase in the
number of long range connections will speed up the recovery towards the steady
state level. It is interesting
to note that an increase of the recovery time with 
increasing $p$ is somewhat contrary to the intuitive idea that
more connections will speed up the time evolution.

The dependence on the mutation probability $p_m$ is more trivial:
The only effect that the mutation probability seems to have
is to speed up the time evolution.
This means that, in the limit of small $p_m$,
the recovery time $\tau$ and the waiting time distribution $P(t_w)$ approaches finite values.  
At $p_m=0.001$ this limit is basically reached for our lattice size
$L=32$. The only effect of a finite $p_m$ in this limit is to prevent
the system from getting stuck in a purely deterministic cycle.

Finally we investigate the case when the influential node is always
defecting. This corresponds to the case when an influential person
does not take any feedback from the environment nor does make any
spontaneous change in its strategy. This does in fact not change any
qualitative features in the behavior of our model.

\section{Conclusions}
We have investigated the spatial Prisoners' Dilemma game for the case
with one influential node. The most striking feature of this model is
the existence of sudden breakdowns of cooperation.~\cite{foot} This is caused by
imitation of a successful scoring by the defector strategy of the influential
node. These breakdowns are associated with two distinct time scales.
One time scale is the recovery time $\tau$ associated with the
recovery back to the steady state cooperation level after a sudden
breakdown. The most interesting feature with this recovery is that it sometimes
becomes slower with increasing small world rewiring. Thus, contrary to the
intuitive feeling that more connections should just speed up the evolution,
it is also possible that the long range connections instead slows down
the time it takes to get back to the equilibrium level. This slowing down of
the recovery occurs when the steady state cooperation level is large enough.  
If the equilibrium cooperation level is small enough
then the recovery time gets shorter with an increasing number of
long range connections.

The second characteristic time is the time between
the sudden breakdowns of cooperation. It is associated with
how often in the steady state situation an event when the
influential node scores highly with the defecting strategy occurs.
This may happen very rarely but when it happens the tendency of the social
network to imitate the influential node causes a dramatic breakdown
of the cooperation level.
The model also contains a random mutation rate.
However this only speeds up the evolution without changing the qualitative behavior.

Our model gives a crude simulation of real social behavior.
However, it does catch a few features of potential interest.
One feature is the instability which an imitating behavior can lead to
in the presence of an influential node be it a charismatic leader,
a popular media person or some such thing. The other is that the restoration
of equilibrium can sometimes be obstructed by the presence of long range
social connections.

One may note that although the present model of asymmetric influence
is quite different in mechanism and spirit from the recent model by
Riolo, Cohen and Axelrod~\cite{axelrod} both
display dynamic instabilities in the cooperation level.

\section*{Acknowledgments}
Comments from Dr.\ F.\ Liljeros are gratefully
acknowledged, as well as financial   
support from the Swedish Natural Research Council through Contract
No.\ F 5102-659/2001.


\begin{thebibliography}{10}
\bibitem{foot} These sudden breakdowns exists also for a random---instead
  of sequential---updating, but are rarer and less dramatic. A quantitative
  study of this case is not within the scope of the present article.
\end{thebibliography}
\end{document}